\documentclass{emulateapj}
\usepackage{graphicx}
\usepackage{amsmath}
\usepackage{units}
\usepackage[normalem]{ulem}
\usepackage{color}
\definecolor{orange}{RGB}{200,100,10}

\newcommand\Teff{T_{\rm eff}}
\newcommand\theff{\theta_{\rm eff}}
\newcommand\TC{T_{\rm C}}
\newcommand\thC{\theta_{\rm C}}
\newcommand\sT{\sigma_{\rm T}}
\newcommand\kB{k_{\rm B}}
\newcommand\tC{t_{\rm C}}
\newcommand\texp{t_{\rm exp}}
\newcommand\ls{\ell_\star}
\newcommand\vs{v_\star}
\newcommand\ths{\theta_\star}
\newcommand\dUC{\dot{U}_{\rm C}}
\newcommand\dUb{\dot{U}_{\rm bulk}}
\newcommand\dUth{\dot{U}_{\rm th}}
\newcommand\dUdamp{\dot{U}_{\rm damp}}
\newcommand\lv{\ell_{\rm damp}}
\newcommand\tv{t_{\rm damp}}

\newcommand\tturb{t_{\rm turb}}
\newcommand\lpl{\ell_{\rm pl}}

\newcommand\fb{f_{\rm bulk}}
\newcommand\fth{f_{\rm th}}
\newcommand\tausw{\tau_{\rm switch}}
\newcommand\V{V}
\newcommand{\figext}[1]{#1.pdf}

\newbox\grsign \setbox\grsign=\hbox{$>$} \newdimen\grdimen \grdimen=\ht\grsign
\newbox\simlessbox \newbox\simgreatbox \newbox\simpropbox
\setbox\simgreatbox=\hbox{\raise.5ex\hbox{$>$}\llap
     {\lower.5ex\hbox{$\sim$}}}\ht1=\grdimen\dp1=0pt
\setbox\simlessbox=\hbox{\raise.5ex\hbox{$<$}\llap
     {\lower.5ex\hbox{$\sim$}}}\ht2=\grdimen\dp2=0pt
\setbox\simpropbox=\hbox{\raise.5ex\hbox{$\propto$}\llap
     {\lower.5ex\hbox{$\sim$}}}\ht2=\grdimen\dp2=0pt
\def\simgt{\mathrel{\copy\simgreatbox}}

\shorttitle{GRB Turbulence}
\shortauthors{J. Zrake, A. M. Beloborodov, C. Lundman}

\begin{document}
\title{Sub-photospheric turbulence as a heating mechanism in gamma-ray bursts}

\author{Jonathan Zrake\altaffilmark{1}, Andrei M. Beloborodov\altaffilmark{1,2} and Christoffer Lundman\altaffilmark{3,4}}

\affil{
$^1$Physics Department and Columbia Astrophysics Laboratory, Columbia University, 538 West 120th St, New York, NY 10027, USA \\
$^2$Max Planck Institute for Astrophysics, Karl-Schwarzschild-Str. 1, D-85741, Garching, Germany \\
$^3$Department of Physics, KTH Royal Institute of Technology, AlbaNova, SE-106 91 Stockholm, Sweden \\
$^4$The Oskar Klein Centre for Cosmoparticle Physics, AlbaNova, SE-106 91 Stockholm, Sweden
}


% =============================================================================
\begin{abstract}
We examine the possible role of turbulence in feeding the emission of gamma-ray bursts (GRBs). Turbulence may develop in a GRB jet as the result of hydrodynamic or current-driven instabilities. The jet carries dense radiation and the turbulence cascade can be damped by Compton drag, passing kinetic fluid energy to photons through scattering. We identify two regimes of turbulence dissipation: (1) ``Viscous'' ---  the turbulence cascade is Compton damped on a scale $\lv$ greater than the photon mean free path $\ls$. Then turbulence energy is passed to photons via bulk Comptonization by smooth shear flows on scale $\ls<\lv$. (2) ``Collisionless'' --- the cascade avoids Compton damping and extends to microscopic plasma scales much smaller than $\ls$. The collisionless dissipation energizes plasma particles, which radiate the received energy; how the dissipated power is partitioned between particles needs further investigation with kinetic simulations. We show that the dissipation regime switches from viscous to collisionless during the jet expansion, at a critical value of the jet optical depth which depends on the amplitude of turbulence. Turbulent GRB jets are expected to emit nonthermal photospheric radiation. Our analysis also suggests revisions of turbulent Comptonization in black hole accretion disks discussed in previous works.
\end{abstract}
% =============================================================================

\maketitle

% =============================================================================
% =============================================================================
\section{Introduction}
\label{sec:introduction}
% =============================================================================

The spectral peak of cosmological gamma-ray bursts (GRBs) is likely emitted at the photosphere of a hot relativistic jet from a compact object (see Beloborodov \& M\'esz\'aros 2017 for a recent review). Four possible dissipation mechanisms have been proposed for the subphotospheric heating of the jet: Compton damping of Alfv\'en waves \citep{Thompson1994}, internal shocks \citep{Rees1994, Levinson2012, Beloborodov2017}, nuclear collisions with free neutrons \citep{Beloborodov2010}, and magnetic reconnection \citep{Drenkhahn2002a}. In this paper we examine heating by strong, fully developed turbulence.

Turbulence is expected to exist in GRB jets on the general grounds that the jet has a large Reynolds number, and must contain inhomogeneities in order to account for the fast variability typically observed in prompt emission. Such non-uniformity of the flow represents a reservoir of mechanical free energy that may be tapped to supply heat. The dissipation of turbulence is expected to occur through a cascade to small scales where the fluid motions are damped. In contrast to jet heating through internal shocks, turbulent heating is volumetric, i.e. distributed in the jet volume. In this respect it is similar to jet heating by collisions between ions and free neutrons.

How and where turbulence is excited depends on the details of jet launching, its interaction with matter around the central engine, magnetization etc. The Kelvin-Helmholtz and Rayleigh-Taylor classes of hydrodynamic instability are examples of mechanisms capable of exciting turbulence \citep{Perucho2005, Duffell2014}. Alternatively, if the jet is magnetically dominated, turbulence can result from magnetic field breakdown via current-driven instabilities such as the kink mode \citep{Bromberg2015}. If the jet contains alternating stripes of magnetic flux analogous to those in pulsar winds \citep{Drenkhahn2002a}, then turbulence may emerge from the breakdown of the stripes \citep{Zrake2016a}.

An attractive feature of turbulence as a heating mechanism is that dissipation occurs on scales smaller than the driving scale of the turbulence cascade, and so the dissipation mechanism may be examined independently of how the cascade is excited. Furthermore, in powerful sources of radiation such as GRBs, the cascade occurs in a dense bath of photons. Then energy may be converted directly from the fluid motions to the radiation field through scattering. Such energy conversion is referred to as bulk Comptonization. Bulk Comptonization may be simpler to calculate than Comptonization by heated/accelerated electrons, since in the latter case the electron energy distribution may be sensitive to uncertain details of collisionless plasma effects.

Photon heating in GRBs by turbulent Alfv\'{e}n waves was proposed earlier by \cite{Thompson1994}. However, this work did not account for wave energy cascading to smaller scales, which might occur faster than its Compton damping. Below we show that the cascade can occur so quickly that Compton damping is rendered inefficient, and then collisionless effects on microscopic plasma scales become responsible for the turbulence dissipation.

Turbulent bulk Comptonization was also studied in the context of black hole accretion disks \citep{Socrates2004, Socrates2006, Kaufman2016}. These works considered cascades extending to the photon mean free path $\ls$, and calculated photon Comptonization by turbulent plasma motions assuming that the plasma is cold. We find that this setup is inconsistent --- as soon as the cascade extends to $\ls$, most of the turbulence energy must be damped into plasma particles and Comptonization by energetic electrons must dominate over bulk Comptonization. We compare our conclusions with the previous work in more detail at the end of the paper.

The paper is organized as follows. In Section~\ref{sec:thermodynamic-evolution} we set up a simplified model for jets heated by turbulence, discuss the energy balance for Comptonization, and define the effective Comptonization temperature. In Section~\ref{sec:photon-heating} we examine how photons gain energy from the turbulence cascade and the effect of Compton drag on the cascade. We identify two regimes of the turbulence damping, viscous and collisionless, and describe the transition from one regime to the other in expanding jets. Finally, in Section~\ref{sec:discussion} we discuss implications of our results for GRBs and accretion disks.

% =============================================================================
% =============================================================================
\section{Heated jets}
\label{sec:thermodynamic-evolution}
% =============================================================================

% =============================================================================
\subsection{Equations of motion}
\label{sec:eom}
We consider the simplest steady model of a radially expanding jet with four-velocity $u^\mu=(\Gamma,\Gamma \V/c,0,0)$ in spherical coordinates $x^\mu=(ct,r,\theta,\phi)$. The jet has proper (baryonic) mass density $\rho(r)$, and its mass flux per unit solid angle is
\begin{equation} \label{eqn:mass}
   \Phi = r^2\rho\,\Gamma \V = const \, .
\end{equation}
The optically thick jet carries trapped radiation. Its stress-energy tensor has the ideal-fluid form,
\begin{equation} 
  T^\mu_\nu = h\rho c^2 u^\mu u_\nu + \delta^\mu_\nu P \, ,
\end{equation}
\begin{equation}
  h = 1 + w, \quad w = \frac{U + P}{\rho c^2} \, .
\end{equation}
Here $w$ is dimensionless enthalpy, $P$ is pressure, and $U=3P$ is internal energy density, which is dominated by radiation.

We will use a toy setup to model turbulent energy injection by tapping into a fictitious external energy reservoir. One could think of this reservoir as another component of the jet that was not included in $T^{\mu\nu}$. The injection is described by adding a source term $\dot{Q} u^\nu$ to the energy-momentum conservation equation,
\begin{equation}
\label{eqn:cons}
	c\,\nabla_\mu T^{\mu \nu} = \dot Q\, u^\nu \, .
\end{equation}
The source is isotropic in the jet rest frame, so that no net momentum is injected in this frame, while the rate of energy injection (which continually stirs turbulence) is given by $\dot Q$. The turbulence energy converts to radiation on the cascade timescale, which is comparable to or shorter than the jet expansion timescale. The main topic of this paper is how the kinetic energy of turbulent motions is passed to photons. However, this section first summarizes the basic thermodynamics of outflows described by Equation~(\ref{eqn:cons}), without considering how the injected energy converts to radiation. Instead, this conversion is assumed to occur immediately, as if $\dot{Q}$ were given directly to photons (which dominate the heat capacity of the jet). 

We will approximate the jet as a steady, uniform conical outflow ($\partial_t = \partial_\theta = \partial_\phi = 0$) with Lorentz factor $\Gamma\gg 1$. In Appendix~A we derive from Equations~(\ref{eqn:mass}) and (\ref{eqn:cons}) two differential equations for $\Gamma(r)$ and $w(r)$, which describe the coupled acceleration and thermal evolution of the jet,
\begin{equation}
\label{eq:Gamma}
 \frac{d\ln \Gamma}{d\ln r}=\frac{2w-\xi(1+w)}{2w+3} \, ,
\end{equation}
\begin{equation}
\label{eq:w}
  \frac{d\ln w}{d\ln r} = \frac{-2w-2+\xi(3w+7+4w^{-1})}{2w+3} \, ,
\end{equation}
where the dimensionless heating parameter $\xi$ is defined by
\begin{equation}
\label{eq:xi}
  \xi=\frac{1}{h\rho c^2}\,\frac{d\,Q}{d\ln r}=\frac{r^3\dot Q}{c^2\Phi h} \, .
\end{equation}
This parameter corresponds to the amount of energy $Q$ that is injected on the jet expansion timescale, compared with the energy existing already in the jet. The expansion timescale (measured in the jet rest frame) is $\texp = r/\Gamma \V$ and the injected energy is $dQ/d\ln r = \dot Q r/\Gamma \V$.

Equations~(\ref{eq:Gamma}) and (\ref{eq:w}) describe jets with any given $\dot{Q}(r)$ or $\xi(r)$. In the model discussed below, $\dot Q$ represents the energy deposition into turbulence by stirring large eddies of size $\ell_0$ and velocity $v_0$. In strong turbulence, the energy density $\sim h\rho v_0^2$ is passed from the largest eddies $\ell_0$ to smaller scales $\ell$ on the eddy turn-over timescale $\sim \ell_0/v_0$. In a quasi-steady cascade, the power flowing
through eddies of scale $\ell$ approximately equals the deposited power $\dot Q \sim h\rho v_0^3/\ell_0$. We assume that the largest scale is comparable to the causally connected scale in the expanding jet, $\ell_0 \sim r/\Gamma$. Then we find
\begin{equation}
\label{eq:xi1}
   \xi \sim \frac{v_0^3}{c^3} \, .
\end{equation}

It is straightforward to integrate Equations~(\ref{eq:Gamma}) and (\ref{eq:w}) for $\Gamma(r)$ and $w(r)$. The solution simplifies in the two limiting cases $w \gg 1$ and $w \ll 1$. For $w\gg 1$, we find in the leading order,
\begin{equation}
 \frac{d\ln \Gamma}{d\ln r}\approx 1-\frac{\xi}{2}, \qquad
 \frac{d\ln w}{d\ln r}\approx -1+\frac{3\xi}{2}
 \qquad
    (w\gg 1).
\end{equation}
Energy injection has a negligible effect if $\xi\ll 1$; then $\Gamma\propto w^{-1}\propto r$.

The opposite regime of $w \ll 1$ occurs at late expansion stages of slowly heated jets, $\xi \ll 1$. In this case, we find (Appendix~A)
\begin{equation}
\label{eqn:wc}
\frac{dw}{d\ln r}\approx -\frac{2}{3}\left(w-2\xi\right)
\qquad (w,\xi\ll 1).
\end{equation}
One can see that $w$ tends to the attractor $w=2\xi$, at which energy injection balances adiabatic cooling. In particular, when $\xi(r)$ is a slowly varying function, one finds a simple solution,
\begin{equation}
  w-2\xi\propto r^{-2/3} \qquad (w,\xi\ll 1).
\end{equation}
Substitution of $w\approx 2\xi$ into Equation~(\ref{eq:Gamma}) yields
\begin{equation}
\label{eq:attr}
   \frac{d\ln \Gamma}{d\ln r}\approx \xi, \qquad
   w\approx 2\xi \qquad (w,\xi\ll 1).
\end{equation}
Without energy injection, $\xi = 0$, this asymptotic state would describe a cold  ballistic flow with $\Gamma=const$ and $w\propto r^{-2/3} \rightarrow 0$ (adiabatic cooling due to $r^2$ expansion). With moderate energy injection, $0 < \xi \ll 1$, the jet cools adiabatically until $w \approx 2\xi$; then $w$ maintains this value while the jet continues to gradually accelerate, $\Gamma \propto r^\xi$.

% =============================================================================
\begin{figure*}
  \includegraphics{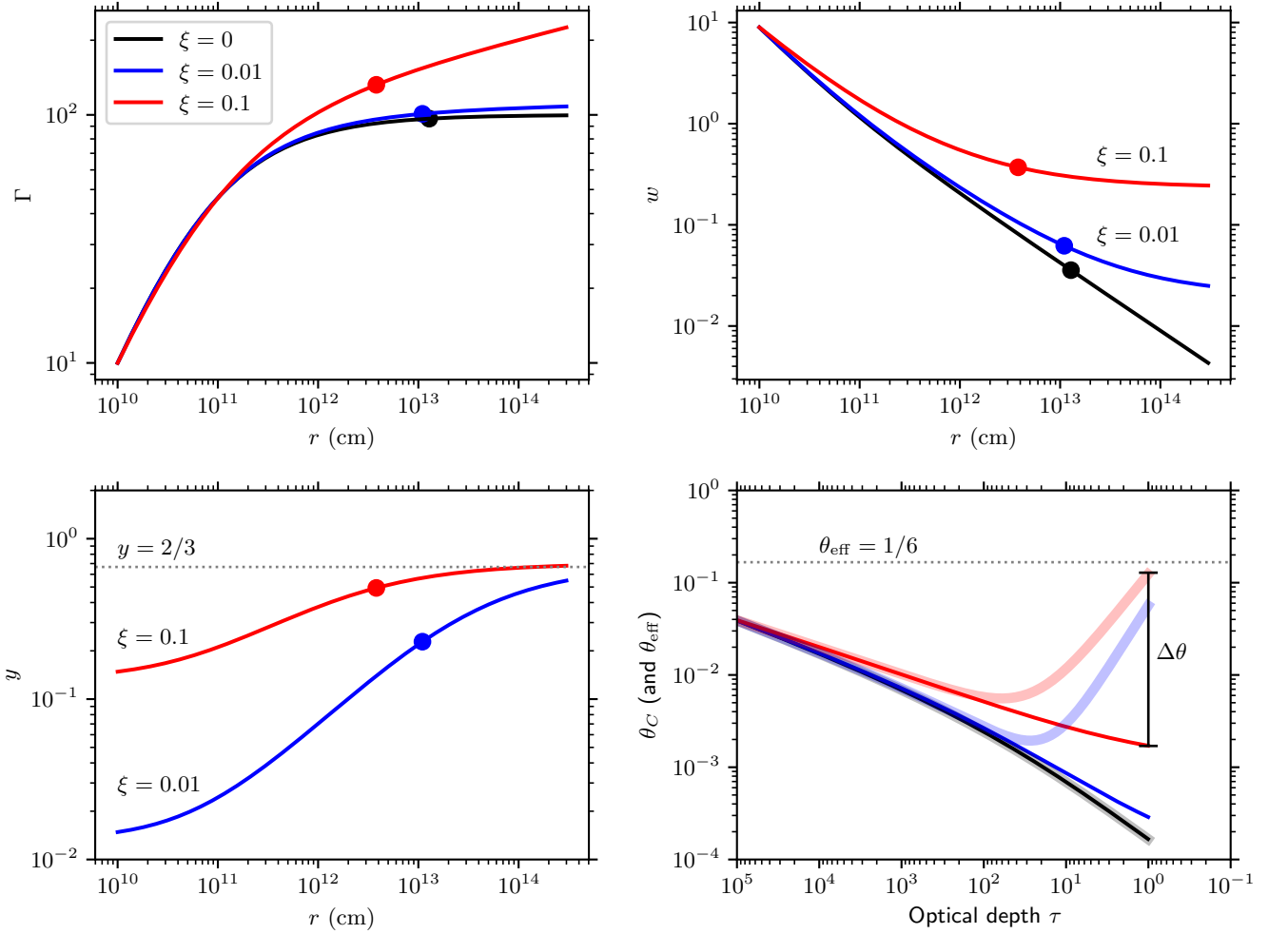}
  \caption{Sample solutions of the jet evolution Equations~(\ref{eq:Gamma}) and (\ref{eq:w}). The solutions describe conical expansion of the jet from $r_1=10^7$~cm, assuming $\Gamma(r_1)=10$, the isotropic equivalent of the jet power $L(r_1)=10^{52}$~erg/s, and $\dot{M}=4\pi\Phi=10^{29}$~g/s. Three models are shown, each with a constant heating parameter $\xi$: $\xi = 0$, $10^{-2}$, and $10^{-1}$. The upper panels show the jet bulk Lorentz factor $\Gamma$ and specific enthalpy $w$. The solid circle on each curve marks the radius where the optical depth of the electron-ion plasma drops to unity. This would be the photosphere of the jet if no $e^\pm$ pairs were created. However, copious pair creation is expected in this region (Section~\ref{sec:discussion}), and therefore the actual photosphere is pushed to a larger radius. The bottom left panel shows the Compton $y$ parameter (Equation~\ref{eq:y}). The bottom right panel shows the approximate Compton temperature $\thC$ assuming photon-to-proton ratio $n_\gamma/n_p=10^5$ and the effective electron temperature $\theff$ from Equation~(\ref{eq:th}) (faint, thick curves). The vertical bar indicates the deviation $\Delta \theta$ of $\theff$ from $\thC$ for the model with $\xi = 0.1$.}
  \label{fig:ThermodynamicEvolution}
\end{figure*}

% =============================================================================
\subsection{Comptonization and the effective electron temperature}
\label{sec:electron-temperature}
%=============================================================================

Radiation receives the injected power $\dot Q$ through the process of Comptonization -- the scattering of photons by randomly moving electrons in the jet frame.\footnote{Other radiative processes typically become important at very high optical depths $\tau \gg 10^4$ \citep[e.g.][]{Beloborodov2013} or in the presence of high-energy particles.} Both thermal and bulk (turbulent) motions contribute to the velocity fluctuations $\delta v_e$ between subsequent scatterings, as described in the next section. Here we evaluate $\delta v_e$ (regardless of its thermal or bulk origin), which may also be associated with an ``effective'' temperature $k\Teff = m_e (\delta v_e)^2/3$. We find it from the energy balance condition, assuming that the injected power $\dot{Q}$ immediately converts to radiation. For simplicity, in this section we imagine that the turbulent fluid is replaced by a static Maxwellian plasma with temperature $\Teff$, although the actual distribution of $\delta v_e$ depends on turbulent motions and may not be Maxwellian.

Let $\dot{U}_{\rm C}$ be the rate of Compton energy exchange between the electrons of density $n_e$ and radiation. It is given by \citep[e.g.][]{Rybicki1979}
\begin{equation}
\label{eq:UC}
  \dUC=4n_e\sT c(\theff-\thC)\,U, \qquad \theff\equiv\frac{\kB \Teff}{m_ec^2}. 
\end{equation}
Here $U$ is radiation energy density and $\thC=\kB\TC/m_ec^2$ is the dimensionless Compton temperature, defined by the condition $\dot U_{\rm C}=0$. At $T=\TC$, the photon energy gain in scattering due to thermal electron motions is balanced by loss due to the Compton recoil effect. $\TC$ is determined by the shape of the radiation spectrum; its typical value in GRB jets is in the keV range.

In GRB jets the heat capacity of electrons is small compared to that of photons, because the ratio of their number densities $n_e/n_\gamma\ll 1$. Furthermore, the timescale for removing the energy from electrons with $\Teff\gg\TC$, $\tC\sim n_e \kB \Teff/\dot{U}_{\rm C}$, is much shorter than the jet expansion timescale,
\begin{equation}
 \tC\ll\texp = \frac{r}{\Gamma c}.
\end{equation}
Electrons moving through the radiation field experience a strong Compton drag force, which does not allow the plasma to store the injected power $\dot Q$; energy is passed immediately to radiation through Compton scattering,
\begin{equation}
   \dUC \approx \dot Q.
\end{equation}
This condition determines the self-regulated $\theff$. It may be re-written in terms of the energy injection parameter $\xi=r\dot Q/\Gamma h\rho c^3$, radiation enthalpy $w=(4/3)U/\rho c^2$,  and the optical depth of the jet $\tau(r)$,
\begin{equation}
\label{eq:th}
    (\theff -\thC)\,\tau\approx\frac{\xi}{3w}\left(1+w\right),
\end{equation}
where
\begin{equation}
\label{eq:tau}
    \tau=\frac{n_e\sT r}{\Gamma}=n_e\sT\ell_0.
\end{equation}

One can see that at large optical depths $\theff \approx \thC$. The decoupling of $\theff$ from $\thC$ occurs at later stages of jet expansion. In particular, for the asymptotic state of a slowly heated jet described by Equation~(\ref{eq:attr}) we find $\theff - \thC \approx (6\tau)^{-1}$. At the photosphere ($\tau=1$), $\theff$ is well above $\thC$ and given by $\theff \approx 1/6$.

The evolution of $\theff$ and $\thC$ is illustrated in Figure~\ref{fig:ThermodynamicEvolution}. We used in this figure a rough approximation\footnote{This approximation would be accurate for radiation with a Wien spectrum.} $\thC \sim \bar{E} / 3 m_e c^2$, where $\bar{E} = U / n_\gamma$ is the average photon energy and $n_\gamma$ is the photon number density. We assumed that the jet carries a constant number of photons per proton, $n_\gamma/n_p=10^5$. One can see that $\thC$ is reduced with increasing radius $r$, because radiation is adiabatically cooled in the expanding jet; $\theff\approx \thC$  follows this cooling trend until $\Delta \theta = \theff - \thC$ grows to high values according to Equation~(\ref{eq:th}).

One can also define the dimensionless Compton parameter,
\begin{equation}
\label{eq:y}
   y\equiv \frac{\dot U_C\texp}{U}=4(\theff-\thC)\tau\approx \frac{4\xi}{3w}\left(1+w\right).
\end{equation} 
Using $w\approx 2\xi\ll 1$ in a slowly heated jet, we find
\begin{equation} 
   y\approx \frac{2}{3}.
\end{equation}
This behavior is observed in the sample models shown in Figure~\ref{fig:ThermodynamicEvolution}. The regime $y \sim 1$ is often called ``unsaturated Comptonization.'' Unsaturated Comptonization with $\theff \gg \thC$ produces nonthermal radiation spectra. 

Thus, turbulence has two effects on the emitted GRB: (1) It prevents strong cooling of radiation in the expanding jet, boosting its photospheric luminosity. (2) As the jet approaches the photosphere, the decoupling of $\theff$ from $\thC$ makes the radiation spectrum increasingly nonthermal.

% =============================================================================
% =============================================================================
\section{Photon heating by turbulence}
\label{sec:photon-heating}

% =============================================================================
\subsection{Sampling of turbulent velocity by photon scattering}
\label{sec:photon-scatting}
% =============================================================================

Photons trapped in the opaque jet ($\tau>1$) are frequently scattered by electrons that move with some velocities $\boldsymbol{v}_e$ in the jet frame. Let $\ls$ be the photon mean free path to scattering. In each scattering event, the photon samples  the distribution of $\boldsymbol{v}_e$ after propagating a distance $\sim\ls$, and this sampling determines the energy gain of the photon --- the process of Comptonization. We wish to understand how turbulent motions contribute to this process. Hereafter the turbulent fluid velocity in the jet frame will be denoted by $\boldsymbol{v}$ and distinguished from the thermal electron velocity $\boldsymbol{v}_{\rm th}$ and the total $\boldsymbol{v}_e=\boldsymbol{v}+\boldsymbol{v}_{\rm th}$. The turbulence spectrum, which describes the amplitudes of turbulent pulsations on various scales $\ell$, will be denoted by $v^2(\ell)$. The turbulence will be assumed isotropic.

Let us first consider the effect of photon Comptonization by turbulence in isolation, i.e. let us assume that the electrons have zero temperature, $T_e=0$, and move only because they are carried by  turbulent eddies relative to the jet frame: $\boldsymbol{v}_e=\boldsymbol{v}$. If the fluid velocity is strongly correlated on scales $\ell\sim\ls$, $\boldsymbol{v}$ would be almost the same in subsequent scattering events, and the photon would not gain energy. The problem would be equivalent to photon scattering by uniformly moving electrons, which becomes trivial if one changes the frame of reference to the rest frame of the correlated electron flow. After a few scatterings radiation is isotropized in this frame, and the photons begin to {\it lose} energy as a result of the Compton recoil effect. 

It is clear that the effect of Comptonization, i.e. upscattering of photons in energy, is controlled by the {\it difference} in $\boldsymbol{v}$ between subsequent scatterings, $\delta \boldsymbol{v}$. Assuming that the random variable $\delta\boldsymbol{v}$ is approximately isotropic, the relevant quantity is the mean expectation of the square of $\delta\boldsymbol{v}$. Let us define
\begin{equation}
  \vs^2\equiv \frac{\overline{(\delta \boldsymbol{v})^2}}{2}.
\end{equation}
Note that $\vs^2$ depends on both the turbulence spectrum and the photon mean free path. In analogy with the corresponding quantity calculated for sampling a Maxwellian distribution, $\vs$ may be used to define an effective ``wave'' temperature, which will characterize the Comptonization effect of turbulence \citep[c.f.][]{Socrates2004,Kaufman2016},
\begin{equation}
\label{eq:the}
   \ths=\frac{\kB T_\star}{m_ec^2}=\frac{\vs^2}{3c^2}.
\end{equation}

Next, let us relax the assumption $T_e=0$. Then one should add to $\boldsymbol{v}$ a random thermal velocity $\boldsymbol{v}_{\rm th}$ drawn from the Maxwellian distribution with temperature $T_e$.  Both $\boldsymbol{v}$ and $\boldsymbol{v}_{\rm th}$ are sampled on the photon free-path scale $\ls$, and we will assume that $T_e\approx const$ on this scale. This gives the net velocity change between subsequent scatterings $\delta\boldsymbol{v}_e = \delta\boldsymbol{v}+\delta\boldsymbol{v}_{\rm th}$, and
\begin{equation}
  \overline{(\delta \boldsymbol{v}_e)^2}=\overline{(\delta \boldsymbol{v})^2} + \overline{(\delta\boldsymbol{v}_{\rm th})^2} 
  = 2\vs^2+\frac{6\kB T_e}{m_e},
\end{equation}
where we have used $\overline{\delta\boldsymbol{v}_{\rm th}\cdot\delta\boldsymbol{v}}=0$ ($\boldsymbol{v}_{\rm th}$ is random and uncorrelated with $\boldsymbol{v}$ or $\delta\boldsymbol{v}$).

The effective temperature discussed in Section~3.1 now becomes
\begin{equation}
\label{eq:thsum}
   \theff=\ths+\theta_e,
\end{equation}
and the Compton power $\dUC$ (Equation~\ref{eq:UC}) can be written as the sum of ``bulk'' and thermal parts,
\begin{eqnarray}
   &\dUC& =\dUb+\dUth, \\
   \label{eq:dUb}
  &\dUb & \approx 4n_e\sT c\,\ths\,U,  \\  
    \label{eq:dUth}
  &\dUth& = 4n_e\sT c\,(\theta_e-\thC)\,U.
\end{eqnarray}
The expression for $\dUb$ is approximate, because the probability distribution for $\delta\boldsymbol{v}$ is not Maxwellian, and so the accurate numerical coefficient in Equation~(\ref{eq:dUb}) may be different from 4. Note also that $\ths$ does not have the meaning of a thermodynamic temperature; it is only a measure of the velocity variations in the turbulent fluid on scale $\ls$. The moving fluid effectively has an infinite mass, and energy is always transferred from turbulent motions to photons, regardless of whether $\ths$ is smaller or larger than the average photon energy. The opposite statement in Socrates et al. (2004) is incorrect. For example, consider the situation with $0<\ths<\thC$ and $\theta_e<\thC$. Then $\dUb>0$  and $\dUth<0$ --- turbulence transfers its energy to photons while the photons are heating the thermal electrons through the Compton recoil effect.

% =============================================================================
\subsection{Radiative damping of the turbulent cascade}
\label{viscous}
% =============================================================================

Photon diffusion with coefficient $D\sim \ls c$ transports energy and momentum through the fluid  and thus creates viscosity.  The corresponding kinematic viscosity coefficient $\nu$ is reduced from $D$ by the ratio of radiation energy density $U$ to the fluid inertial mass density $h\rho c^2$, and so
\begin{equation}
\label{eq:nu}
   \nu\sim \frac{w}{1+w}\,\ls c,
\end{equation}
where we used $U/h\rho c^2=(3/4)(w/h)$ and $h=1+w$. The radiation viscosity in GRB jets dominates over viscosity due to transport of plasma particles, because their free paths are much shorter than $\ls$ and they are slower than photons.

The viscosity coefficient $\nu$ determines the Reynolds number of turbulence stirred on scale $\ell_0\sim r/\Gamma$ with velocity $v_0$,
\begin{equation}
\label{eq:Re}
   Re=\frac{\ell_0 v_0}{\nu}\sim \frac{1+w}{w}\,\tau\,\frac{v_0}{c}\sim \frac{1+w}{w}\,\tau\,\xi^{1/3}.
\end{equation}
Normally, viscosity is expected to cut off the turbulence cascade at scales $\ell$ smaller than some damping scale $\lv$. Turbulent motions with $\ell\gg\lv$ are not affected by viscous dissipation, and the cascade has the standard Kolmogorov spectrum,
\begin{equation}\label{eq:cascade-spectrum}
  v(\ell) \equiv\left(\frac{dv^2}{d\ln\ell}\right)^{1/2} = v_0 \left(\frac{\ell}{\ell_0}\right)^{1/3}, \qquad \ell\gg\lv.
\end{equation}
The rate of viscous damping of turbulent eddies of scale $\ell$ is given by
\begin{equation}
  \dUdamp(\ell)\sim \frac{h\rho v^2(\ell)}{\tv(\ell)}, \qquad \tv(\ell)\sim \frac{\ell^2}{\nu}.
\end{equation}
Note that $\dUdamp(\ell)$ scales with $\ell$ as $\ell^{-4/3}$. In the inertial range $\ell\gg\lv$ the cascade power $\dot{Q}$ moves from larger to smaller scales without significant losses, $ \dUdamp(\ell)\ll\dot{Q}$. The viscous cutoff occurs at $\lv$ where $\dUdamp$ becomes comparable to $\dot{Q}$, which gives
\begin{equation}
\label{eq:lv}
  \lv\sim \ell_0\, Re^{-3/4}\sim \ls \tau \, Re^{-3/4} \qquad (\lv>\ls).
\end{equation}
This expression is meaningful only if $\lv$ is larger than the photon mean free path $\ls$, so that the diffusion picture of photon transport is valid on all scales $\ell\simgt\lv$. One can see from Equation~(\ref{eq:lv}) that
\begin{equation}
  \lv\gg\ls \quad \Leftrightarrow \quad Re\ll\tau^{4/3}.
\end{equation}
Hereafter the regime $Re<\tau^{4/3}$ is called ``viscous.'' The opposite regime will be discussed in Section~\ref{sec:collisionless}.

The condition $\ls<\lv$ implies that photon scattering samples $\delta\boldsymbol{v}$ on scales where turbulent fluctuations are suppressed by Compton drag. Then $\vs^2=\overline{(\delta\boldsymbol{v})^2}$/2 may be estimated as follows. A velocity field that varies by $v(\lv)\sim v_0(\lv/\ell_0)^{1/3}$ on scale $\lv$ and is smooth on scales $\ell<\lv$ can be described as a shear flow with $|\nabla v|\sim v(\lv)/\lv$. Fluid compression is negligible on small scales, because $v(\lv)$ is far subsonic.
% \cite[including for relativistic turbulence][]{Zrake2012})
% on these scales (and so the flow is nearly pure shear) because $v(\lv)$ is far subsonic.
The velocity variation on scale $\ls$ is then given by
\begin{equation}
\label{eq:vs_viscous}
   \vs\approx \frac{\ls}{\lv}\,
   v(\lv) =\frac{v_0}{\tau}\,\sqrt{Re}    \quad (Re\ll \tau^{4/3}),
\end{equation}
where we used the relation $\tau\approx\ell_0/\ls$ for $\ell_0\approx r/\Gamma$.

Substitution of Equation~(\ref{eq:Re}) for $Re$ yields the wave temperature (Equation~\ref{eq:the}),
\begin{equation}
   \ths\approx \frac{1+w}{w}\,\frac{\xi}{3\tau}  \qquad (Re \ll \tau^{4/3}) \, .
\end{equation}
It is easy to verify that this $\ths$ gives $\dUb$ (Equation~\ref{eq:dUb}) equal to $\dot{Q}$, as expected in the viscous regime $Re\ll\tau^{4/3}$ --- almost all the cascade power is passed to photons via bulk Comptonization. It is also consistent with Equations~(\ref{eq:th}) and (\ref{eq:thsum}).

We conclude that, in the viscous regime, Comptonization of radiation is powered by the smooth velocity shear rather than a random turbulent velocity field. The shear is a residual smooth variation of the fluid velocity after damping the turbulence on all scales $\ell<\lv$. Thus, the process of photon heating by turbulence may be called ``shear Comptonization'' more accurately than ``turbulent bulk Comptonization.''

To avoid any confusion, note that the fluid velocity variation $\delta \boldsymbol{v}(\ell)$ on a scale $\ell$ scales linearly with $\ell$ in the viscous sub-range ($\ell < \ell_{\rm damp}$) even though the cascade spectrum $v(\ell)$ is strongly suppressed at $\ell<\lv$. This is consistent with the general relationship between the two-point correlation function and the power spectrum. For instance, consider a function $f(x)=\sin (x/a)$: it varies linearly on scales much smaller than $a$ while its Fourier spectrum $f_k=0$ for all $k\neq a^{-1}$.

% =============================================================================
\subsection{Collisionless dissipation regime}
\label{sec:collisionless}
% =============================================================================

Next, let us consider the regime $Re>\tau^{4/3}$. In this case, it turns out that Compton drag fails to damp turbulent motions on any scale $\ell$.

The difference between the regimes $Re<\tau^{4/3}$ and $Re>\tau^{4/3}$ is illustrated in Figure~2. In the former regime, the power law $v(\ell)=v_0(\ell/\ell_0)^{1/3}$ cuts off at $\lv$. By contrast, when $Re>\tau^{4/3}$ the cascade proceeds unimpeded by Compton drag. This fact can be understood by looking at two relevant timescales --- the timescale for radiative damping of an eddy, $\tv(\ell)$, and the cascade timescale $\tturb(\ell)=\ell/v(\ell)$. The damping effect is small if $\tv>\tturb$. 

Note that $\tv(\ell)$ is independent of the eddy's speed $v(\ell)$, because its energy and radiative losses are both proportional to $v^2(\ell)$. The damping timescale is given by
\begin{equation}
   \tv(\ell) \sim \frac{1+w}{w}\,\frac{\ls}{c}\left\{\begin{array}{ll}
        (\ell/\ls)^2 & \quad \ell>\ls \\
        1              & \quad \ell<\ls
                                                                      \end{array}
                                                              \right.
\end{equation}
Damping on scales $\ell>\ls$ occurs in the diffusion regime, and then $\tv\propto\ell^2$. On these large scales radiation is trapped and advected by the turbulent flow, and photon transport through the fluid (causing viscosity) is a relatively small effect described in the diffusion approximation.

Damping at $\ell<\ls$ is different, because on the small scales photons  stream freely through the fluid. The smallest eddies still capable of trapping radiation have sizes $\sim\ls$. On smaller scales radiation is approximately uniform and also quasi-isotropic, when viewed in the local frame comoving with the eddy $\ell\sim\ls$. In this frame, the smaller superimposed eddies $\ell<\ls$ move through the approximately isotropic radiation, and their energy losses are described by the standard Compton drag formula $\dUdamp\approx (4/3)\sT c\, n_e U(v/c)^2$. It gives the damping timescale $\tv\sim h\rho v^2/\dUdamp\sim(h/w)(\ls/c)$.

% =============================================================================
% =============================================================================
\begin{figure}[t]
\includegraphics[width=0.51\textwidth]{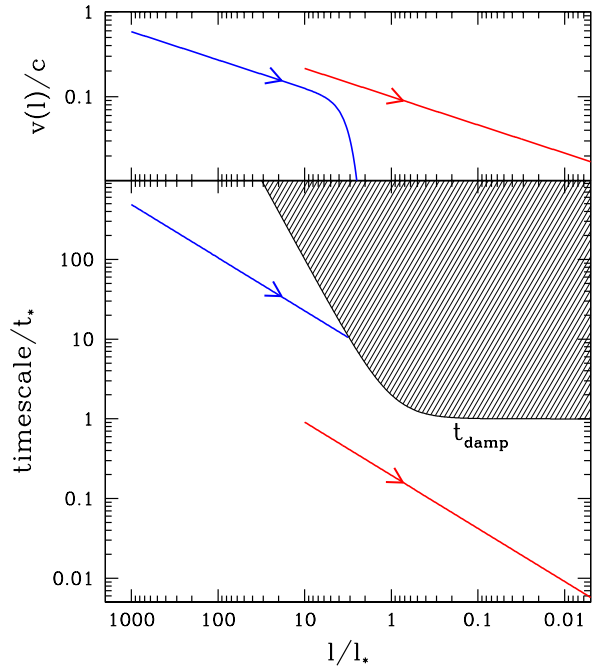}
\caption{Upper panel: Turbulence cascade spectrum $v(\ell)$ in the presence of Compton damping. Blue curve shows the cascade in a jet with optical depth $\tau=\ell_0/\ls=10^3$ and the heating parameter $\xi=0.2$; it has a viscous cutoff at $\lv>\ls$. The red curve shows a similar cascade when $\tau=10$ and $\xi=0.01$; it does not cut off and extends to microscopic plasma scales $\ell\ll\ls$. In both cases, the jet has enthalpy $w\approx 2\xi$ (see text). Bottom panel: comparison of the cascade timescale $t_{\rm turb}(\ell)=\ell/v(\ell)$ (blue and red curves show the same models as in the upper panel) with the Compton damping timescale $\tv(\ell)$ (black curve). Compton damping prevents the cascade from entering the shaded region where $t_{\rm turb}>\tv$.}
\label{fig:cascade}
\end{figure}
% =============================================================================

At all $\ell$ where damping is negligible, the cascade has the Kolmogorov spectrum 
and its characteristic timescale (the time for passing energy to smaller eddies) is comparable to the eddy turnover time,
\begin{equation}
   \tturb(\ell)=\frac{\ell}{v}=\frac{\ell_0}{v_0}\left(\frac{\ell}{\ell_0}\right)^{2/3}.
\end{equation}
The condition $Re>\tau^{4/3}$ is equivalent to $\tturb(\ls)<\tv(\ls)$. Then $\tturb(\ell)<\tv(\ell)$ at all $\ell$, as one can see in Figure~2. Thus, the cascade power continues to flow toward smaller scales unimpeded by Compton drag. The cascade proceeds with the Kolmogorov spectrum to a microscopic scale $\lpl\ll\ls$ where damping eventually occurs through collisionless plasma effects. As a result, almost all the power $\dot{Q}$ is deposited into the plasma particles rather than directly passed to photons via bulk Comptonization.

The reduced efficiency of bulk Comptonization in this regime can be estimated as follows. In contrast to the viscous regime, photon scattering samples the random velocities of fully developed turbulent eddies $\ell\sim\ls$ with
\begin{equation}
\label{eq:vs2}
   \vs\approx v_0\left(\frac{\ls}{\ell_0}\right)^{1/3}=\frac{v_0}{\tau^{1/3}} \qquad (Re\gg\tau^{4/3}),
\end{equation}
which gives
\begin{equation}
\label{eq:ths2}
  \ths\approx \frac{1}{3}\left(\frac{\xi}{\tau}\right)^{2/3}  \qquad (Re\gg\tau^{4/3}),
\end{equation}
where we used the relation $\xi\approx v_0^3/c^3$ (Equation~\ref{eq:xi1}). 
The net bulk Comptonization power $\dUb$ (Equation~\ref{eq:dUb}) approximately equals the energy losses of eddies $\ell\sim\ls$, $\dUb\approx \dUdamp(\ls)$. Using Equation~(\ref{eq:ths2}) it is easy to find the ratio of $\dUb$ to the total cascade power $\dot{Q}$, 
\begin{equation}
\label{eq:fb}
    \fb\equiv\frac{\dUb}{\dot{Q}}\approx \frac{w}{1+w}\left(\frac{\tau}{\xi}\right)^{1/3}=\frac{\tau^{4/3}}{Re}\ll 1.
\end{equation}
The small $\fb$ is consistent with the cascade delivering the larger part of its power, $\dot{Q}-\dUb$, to the plasma scales and dissipating there. The dissipated power  $\dot{Q}-\dUb$ eventually converts to radiation through cooling of the heated plasma particles. In particular, if this power is received and radiated by thermal (Maxwellian) electrons, then $\dot{Q}-\dUb=\dUth$ is described by Equation~(\ref{eq:dUth}), and the thermal fraction of the total Comptonization power is 
\begin{equation}
\label{eq:fth}
   \fth=\frac{\dUth}{\dot{Q}}=1-\fb.
\end{equation}
In more realistic models, the electron distribution created by collisionless dissipation may be nonthermal, not described by the single parameter $\theta_e$. Furthermore, the inverse Compton emission from nonthermal particles could be accompanied by synchrotron emission. However, the total power radiated by the heated/accelerated particles still equals $(1-\fb)\dot{Q}$. The fraction $f_{\rm bulk}$ vanishes in the limit $\tau \ll 1$. Then all turbulent energy converts to energetic particles, which radiate through inverse Compton and synchrotron cooling. This situation was recently discussed by \cite{Uzdensky2018}.

% =============================================================================
\subsection{Switch from Compton damping to collisionless dissipation}
\label{sec:switch}
% =============================================================================

An expanding GRB jet initially has a huge optical depth $\tau$ and turbulence dissipates in the viscous regime $Re\ll \tau^{4/3}$. The entire power of the cascade is passed to radiation via bulk (shear) Comptonization as described in Section~3.2, so that $\fb=1$ and $\fth=0$. 

Collisionless dissipation of turbulence on small scales switches on when $\tau^{4/3}/Re=(\tau/\xi)^{1/3} w/(1+w)$ drops below unity, as described by Equation~(\ref{eq:fb}). Thus the switch from Compton damping to collisionless dissipation occurs at the characteristic optical depth
\begin{equation}
   \tau_{\rm switch}\sim\xi\, \frac{(1+w)^3}{w^3}\sim \frac{\xi}{w^3}, 
\end{equation}
where the last equality assumes that the jet has cooled to $w<1$ before reaching $\tau_{\rm switch}$. In particular, in the asymptotic state of the slowly heated jet (Equation~\ref{eq:attr}) we find
\begin{equation}
   \tau_{\rm switch}\sim \frac{1}{8\,\xi^2}. 
\end{equation}
Then the history of bulk Comptonization in the expanding jet may be summarized as follows
\begin{equation}
  \fb\approx\left\{\begin{array}{ll}
       1 & \quad \tau>\tau_{\rm switch} \\
       (\tau/\tau_{\rm switch})^{1/3} & \quad \tau<\tau_{\rm switch}
                          \end{array}
                  \right.
\end{equation}
Collisionless dissipation is negligible at $\tau>\tau_{\rm switch}$ and gradually increases at smaller optical depths. In particular, at the photosphere $\tau=1$ we find $\fb\approx 2\xi^{2/3}$ and $\fth\approx 1-2\xi^{2/3}$.

% =============================================================================
\section{Discussion}
\label{sec:discussion}
% =============================================================================

% =============================================================================
\subsection{Compton damping of turbulence}

In this paper we have examined how turbulence energy converts to radiation through Compton drag. We focused on GRB jets, however our analysis can also be applied to other objects, in particular accretion disks around black holes (see below). Our results may be summarized as follows.

We have shown that turbulence passes its energy directly to radiation (via bulk Comptonization of photons) if the following condition is satisfied,
\begin{equation}
\label{eq:cond}
   \tau>
  % \frac{v_0}{c}} \left(\frac{w}{1+w}\right)^{-3}  
   \left(\frac{v_0}{c}\,\frac{1+w}{w}\right)^{3}  
  \quad {\rm (efficient~Compton~damping)}. 
\end{equation}
Here $v_0$ is the speed of the largest turbulent eddies on the driving scale $\ell_0$, $\tau=\ell_0/\ls$ is the optical depth of the largest eddies (comparable to the optical depth of the jet), and $w=(4/3)U/\rho c^2$ is the dimensionless enthalpy of radiation. 
We called this regime ``viscous,'' because the condition~(\ref{eq:cond}) implies that photon viscosity cuts off the turbulence
cascade. 
The condition~(\ref{eq:cond}) may also be formulated in terms of the turbulence Reynolds number $Re$ as $\tau>Re^{3/4}$ (Section~3).
In the opposite regime, when $\tau$ is below the critical value, the cascade proceeds to microscopic plasma scales, losing only a fraction of its power to Compton drag (Equation~\ref{eq:fb}).

In GRB jets, it is convenient to describe turbulence power using the dimensionless parameter $\xi$ --- the energy given to turbulence on the jet expansion timescale, normalized to the bulk kinetic power of the jet. We have shown that the enthalpy of heated jets tends to $w\approx 2\xi$ (Section~2). Using this relation and the approximate expression $v_0/c\sim\xi^{1/3}$ (for strong turbulence), we estimated the critical optical depth $\tausw\sim (8\xi^2)^{-1}$. When the optical depth of the jet drops below $\tausw$, Compton drag becomes inefficient and turbulence dissipation switches from the viscous to collisionless regime.

We have also argued that when Compton damping is efficient (viscous regime) photons gain energy from smooth shear layers, whose velocity profiles are shaped by radiation viscosity. This dissipation mechanism resembles radiation-mediated shocks, although in a shock photons gain energy from a converging bulk flow, rather than shear.

One aspect of turbulent heating was not explored in our paper: turbulence is generally intermittent, which implies disproportionate concentration in space and time of the energy dissipation \citep[e.g.][]{She1994}. Therefore, its Compton damping is not uniform. In the viscous regime, the shearing layers within which photons are most vigorously heated, will occupy only a fraction of the jet volume. The ``wave temperature'' $\ths$ describing bulk Comptonization by fluid motions and contributing to $\theff=\theta_e+\ths$ (Section~\ref{sec:photon-heating}) represents only an average effect of turbulence on photons. Sampling of the intermittent velocity field (as photons sample in each scattering event) may be quite different from sampling a Gaussian distribution.

% =============================================================================
\subsection{GRB radiation from subphotospheric turbulence}

The evolution of radiation carried by heated and opaque GRB jets proceeds through three radial zones (Beloborodov 2013): blackbody zone ($\tau\simgt 10^5$) where radiation is Planckian, Wien zone ($\tau\simgt 10^2$) where photons and electrons still maintain kinetic equilibrium at a common temperature $T_e=T_{\rm rad}=T$ but $n_\gamma$ may be below its blackbody value $aT^3/2.7k$, and nonthermal zone ($\tau<10^2$) where $T_e>T_{\rm rad}\sim T_{\rm C}$. The decoupling of $T_e$ from $\TC$ occurs at optical depths $\tau<\Theta_{\rm C}^{-1}\sim 10^2$ (Figure~1) and leads to Comptonization of photons that broadens their spectrum and creates a high-energy tail (see radiative transfer simulations in Beloborodov 2010; Vurm \& Beloborodov 2016). 

Radiation spectra in jets heated by turbulence will follow this general evolution and emit a nonthermal spectrum at the photosphere, with a peak at 0.1-1~MeV. Given the parameter $\xi$ that describes the power injected in turbulence, the generated radiation spectrum could be determined from first principles. However, there are a few factors that  complicate the exact spectrum calculations and require future work:

(1) Copious $e^\pm$ pair creation becomes inevitable if the turbulent heating extends to moderate optical depths $\tau\simgt 1$, where the effective plasma temperature $\theff$ exceeds $\sim 0.1$ according to the approximate relation $y\equiv 4(\theff-\thC)\tau\approx 2/3$ (Section~2). The created $e^\pm$ will increase the plasma opacity and push the photosphere to a larger radius. Nevertheless, the energy balance condition $y\approx 2/3$ will still hold, and hence the steep increase of $e^\pm$ abundance with the growing $\theff\approx (6\tau)^{-1}$ will buffer the temperature growth. As a result, $\tau(r)$ will keep a moderate value $\tau\simgt 1$ in an extended range of radii while $\theff$ slowly grows from $\sim 0.1$ toward $\sim 1$. This radial ``stretching'' of the photospheric layers $\tau\simgt 1$ will affect the Comptonized spectrum emerging from the jet.

(2) In the collisionless dissipation regime ($\tau<\tausw$), it is presently unknown how the dissipated power is partitioned between the electrons/positrons and the ions. Ions do not radiate their energy and may be unable to pass it to $e^\pm$ via Coulomb collisions at optical depths $\tau<20$ (Beloborodov 2010).

(3) Collisionless dissipation of turbulence may lead to generation of nonthermal electrons \citep{Zhdankin2018}. This would impact the emitted photospheric spectrum.  Inverse Compton emission from the accelerated particles will lead to secondary generations of pair creation. Furthermore, the energetic $e^\pm$ produce significant synchrotron radiation, avoiding self-absorption. Radiative transfer simulations (Vurm \& Beloborodov 2016) show that nonthermal particles have a strong effect on the photospheric spectrum emitted by GRB jets, and make it more consistent with observations.

% =============================================================================
\subsection{Comparison with previous work on turbulent Comptonization}
\label{sec:socrates}
% =============================================================================

\cite{Thompson1994} considered Alfv\'en waves generated by magnetic reconnection events inside an opaque, magnetically dominated jet. His picture assumed that the waves pass their energy to the radiation field through bulk Comptonization with an effective Compton parameter $y \sim 1$, and this condition was used to determine the relation between the effective wave temperature and the jet optical depth. This picture neglected the cascade of the wave energy to small scales, and the fact that the correct effective wave temperature for bulk Comptonization $\ths$ is set by the speed of turbulent motions on the small scale $\ls$ (the photon mean free path). \cite{Thompson1994} did not consider the possibility that bulk Comptonization consumes only a fraction of the turbulence power, with the remaining power being dissipated on much smaller scales by collisionless plasma processes (Section \ref{sec:collisionless}). The switching from viscous to collisionless dissipation regimes at $\tau<\tausw$ (Section \ref{sec:switch}) was not recognized.

The dependence of wave temperature $\ths$ on the scale $\ls$ was correctly defined in \cite{Socrates2004} and also in \cite{Kaufman2016}. Both of these studies focused on the role of turbulent Comptonization in shaping the spectrum of coronal X-ray emission from black hole accretion disks. \cite{Kaufman2016} computed in detail $\ths$ for different turbulent energy spectra and studied the influence of varying levels of compressive versus vortical turbulent motions.
These studies focused on the regime where bulk Comptonization by turbulence dominates over thermal Comptonization, which is equivalent to the viscous regime in our terminology. At the same time, they assumed that the turbulence cascade extends to the scale $\ls$. This assumption is inconsistent, because in the viscous regime the cascade is cut off on a scale $\lv>\ls$ (Section~3).

In the regime where the cascade does extend to scale $\ls$, it will not be stopped by Compton drag on any scale (Figure~2) and most of the turbulence power will be deposited into the plasma particles (the collisionless dissipation regime). Then it is inconsistent to assume a cold plasma, as assumed in the simulations of turbulent Comptonization in \cite{Socrates2004}.

We also point out that contrary to the expectation of \cite{Socrates2004} bulk Comptonization can dominate even when $\ths<\theta_e$. This occurs at large optical depths $\tau>\thC^{-1}$, as explained in Section~3.1: the electron temperature $\theta_e$ stays close to the Compton temperature $\thC$ (thermal Comptonization vanishes $\theta_e=\thC$) while bulk Comptonization with a small $\ths$ gradually heats the photons.

% =============================================================================
\subsection{Compton damping of turbulence in accretion disks}
\label{sec:accretion-disks}
% =============================================================================

Accretion disks are turbulent, and dissipation of the turbulence cascade can occur through Compton damping \citep{Socrates2004, Kaufman2016, Kaufman2018}. Our results imply that turbulence in accretion disks may be dissipated in the two different regimes, viscous and collisionless, depending on the parameters of the disk, in particular its accretion rate $\dot M$.

Accretion disks have dimensionless enthalpy $w<1$, and the critical optical depth for switching to collisionless dissipation (Equation~\ref{eq:cond}) simplifies to
\begin{equation}
   \tau_{\rm switch}\sim \left(\frac{v_0/c}{w}\right)^3. 
\end{equation}
We now estimate $\tau_{\rm switch}$ for the standard $\alpha$-disk model \citep{Shakura1973}. We are particularly interested in radiation-dominated disks around black holes with sufficiently high accretion rates in the Eddington units,
\begin{equation}
  \dot{m}=\frac{\dot{M} c^2}{L_E},
  \qquad L_E=\frac{2\pi r_g c^3}{\kappa},
\end{equation}
where $\kappa$ is the plasma opacity and $r_g=2GM/c^2$ is the gravitational radius of the black hole. Then the disk thickness $2H$ is set by the local radiation flux at the disk surface,

\begin{equation}
  F = \frac{3}{8}\, \frac{GM\dot{M}}{r^3}\,S, \qquad S(r)=1-\left(\frac{r_{\rm in}}{r}\right)^{1/2},
\end{equation}
where $r_{\rm in}$ is the inner radius of the disk (the last stable Keplerian orbit; for a Schwarzschild black hole $r_{\rm in}=3r_g$). The disk thickness is found from the vertical force balance $F\kappa/c=GMm_p H/r^3$, which gives
\begin{equation}
 H\approx \frac{3}{8\pi} \frac{\kappa\dot{M}}{c}\,S= \frac{3}{4}\,\dot{m}\,S\,r_g.
\end{equation}
The approximate energy density inside the disk $U\sim \rho H\kappa \,F/c$ gives an estimate for the dimensionless enthalpy $w=(4/3)U/\rho c^2$,
\begin{equation}
   w\sim \frac{HF\kappa}{c^3}\sim \frac{\dot{m}^2S^2}{x^3}, \qquad x\equiv \frac{r}{r_g}.
\end{equation}
The turbulent speed of the largest eddies satisfies the approximate relation,
\begin{equation}
   \frac{v_0^2}{c^2}\sim \chi w, \qquad \chi\equiv \frac{U_{\rm turb}}{U},
\end{equation}
where $U_{\rm turb}$ is the turbulence energy density. This gives
\begin{equation}
   \tau_{\rm switch}\sim 
   \left(\frac{\chi}{w}\right)^{3/2}
   \sim \frac{\chi^{3/2} x^{9/2}}{\dot{m}^3S^3}. 
\end{equation}

The largest turbulent eddies have size $l_0 < H$, and the eddy optical depth is given by
\begin{equation}
  \tau\sim \kappa\rho\, l_0\sim \frac{x^{3/2}}{\alpha\,\dot{m}\,S}\,\frac{\ell_0}{H}.
\end{equation}
Here we used the relation $\rho\approx \dot{M}/4\pi r H v_{\rm acc}$ and the standard expression for the accretion speed in an $\alpha$-disk, $v_{\rm acc}=\alpha\,S^{-1} (H/r)^2 (GM/r)^{1/2}$ \citep{Shakura1973}.

Comparing $\tau$ and $\tau_{\rm switch}$ we conclude that turbulence will be Compton damped in disks with sufficiently high accretion rates $\dot{m}>\dot{m}_1$, where
\begin{equation}
  \dot{m}_1\sim \alpha^{1/2}\left(\frac{U_{\rm turb}}{U}\right)^{3/4} \left(\frac{\ell_0}{H}\right)^{-1/2}\,
  \frac{r/r_g}{1-(r_{\rm in}/r)^{1/2}}.
\end{equation}
Turbulence in radiation-dominated disks with $\dot{m}<\dot{m}_1$ should have a fully developed cascade down to microscopic plasma scales, where it is dissipated by collisionless effects. One can see that the opposite, Compton drag-dominated, regime occurs in the inner regions of very bright accretion disks, approaching or exceeding the Eddington limit.

Note that $m_1$ depends on radius $r$, and for a given accretion rate one can define a characteristic radius $r_1$ inside of which $\tau>\tau_{\rm switch}$. In the case of super-Eddington accretion, it may be useful to compare $r_1$ with the ``trapping'' radius $r_{\rm tr}\approx\dot{m} r_g S^{1/2}$ where the accretion flow begins to trap and advect radiation, instead of emitting it. Both $r_1$ and $r_{\rm tr}$ scale as $\dot{m}$. The ratio
\begin{equation}
  \frac{r_1}{r_{\rm tr}}\sim \alpha^{-1/2}\left(\frac{U_{\rm turb}}{U}\right)^{-3/4} \left(\frac{\ell_0}{H}\right)^{1/2}
  \qquad (\dot{m}\gg 1),
\end{equation}
is expected to exceed unity. Thus, Compton damping of turbulence should generally be important in super-Eddington accretion disks.

The $\alpha$-viscosity model used in the above estimates is incomplete, and significant efforts have recently been invested into numerical simulations of radiation-dominated disks, including the super-Eddington regime \citep[e.g.][]{Jiang2014, Jiang2017}. The effects discussed above, in particular the switch from viscous to collisionless dissipation can be observed in the simulations as long as they resolve scales smaller than the photon mean free path.

\acknowledgements
A.M.B. is supported by NSF grant AST-1816484, NASA grant NNX15AE26G, and a
grant from the Simons Foundation \#446228.

\appendix

For a steady, uniform conical flow ($\partial_t=\partial_\theta = \partial_\phi = 0$), one finds from the energy and momentum conservation laws ($t$ and $r$ components of Equation~(\ref{eqn:cons})),
\begin{eqnarray}
 \frac{c^2}{r^2}\,\frac{d}{dr} (\Phi c^2 h\Gamma) &=& \dot Q \,\Gamma 
\label{eqn:energy}\, , \\
 \frac{c^2}{r^2}\,\frac{d}{dr} (\Phi h \Gamma v) + c^2\,\frac{dP}{dr} &=& \dot Q\,\Gamma \V
\label{eqn:momentum} \, .
\end{eqnarray}
Combining Equations~(\ref{eqn:energy}) and (\ref{eqn:momentum}) one finds
\begin{equation}
\label{eqn:acc}
  \frac{dv}{dr}=-\frac{r^2}{\Phi h\Gamma}\,\frac{dP}{dr}.
\end{equation}
$\dot Q$ dropped out from this equation, which simply states that the flow is accelerated by the pressure gradient force.

One can express Equations~(\ref{eqn:energy}) and (\ref{eqn:acc}) as two differential equations for two unknowns, e.g. $u\equiv\Gamma\, \V/c=\Gamma\beta$ and $w$. Equation~(\ref{eqn:energy}) gives (using $dh=dw$ and $d\Gamma=\beta du$),
\begin{equation}
\label{eqn:xi1}
  \frac{dw}{d\ln r}=(1+w)\left(\xi- \beta^2\,\frac{d\ln u}{d\ln r}\right), \qquad \xi\equiv\frac{r^3\dot Q}{c^2\Phi h}.
\end{equation}
Equation~(\ref{eqn:acc}) can be rewritten using $P=w\rho c^2/4=w\Phi c/4ur^2$,
\begin{equation}
\label{eqn:acc1}
  \frac{d\V}{dr}=-\frac{r^2}{4ch\Gamma}\,\frac{d}{dr}\left(\frac{w}{r^2u}\right).
\end{equation}
Then substituting $d\V=c\,du/\Gamma^3$ one finds
\begin{equation}
\label{eqn:acc2}
  \left(\frac{4}{3}\,{\cal M}^2-1\right)\frac{d\ln u}{d\ln r} =2-\frac{d\ln w}{d\ln r},
  \qquad
  {\cal M}^2 \equiv \frac{\V^2}{c_s^2}=\frac{3(1+w)}{w}\,\frac{\V^2}{c^2},
\end{equation}
where $c_s^2=4P/3h\rho=c^2w/3h$ is the local sound speed.

Equations~(\ref{eqn:xi1}) and (\ref{eqn:acc2}) give two differential equations for $u$ and $w$,
\begin{equation}
\label{eqn:u}
  \left({\cal M}^2-1\right)\frac{d\ln u}{d\ln r}=2-\frac{\xi}{w}(1+w),
\end{equation}
\begin{equation}
\label{eqn:w}
 \frac{d\ln w}{d\ln r}=2-\frac{(4{\cal M}^2/3-1)}{({\cal M}^2-1)}\left[2-\frac{\xi}{w}(1+w)\right].
\end{equation}
Equations~(\ref{eqn:u}) and (\ref{eqn:w}) describe a conical, steady, relativistic outflow with any energy injection profile $\dot Q(r)$ or $\xi(r)$. Note that the flow accelerates only when it is supersonic. 

GRB jets are ultrarelativistic, and we will consider flows with $u\approx\Gamma\gg 1$. Then Equations~(\ref{eqn:u}) and (\ref{eqn:w}) simplify to
\begin{equation}
\label{eqn:u1}
 \frac{d\ln \Gamma}{d\ln r}=\frac{2w-\xi(1+w)}{2w+3},
\end{equation}
\begin{equation}
  \label{eqn:w1}
  \frac{d\ln w}{d\ln r} = \frac{-2w-2+\xi(3w+7+4/w)}{2w+3} 
 = -\frac{(2-3\xi)(w-w_1)(w+1)}{w(2w+3)},
 \qquad
  w_1=\frac{4\xi}{2-3\xi}.
\end{equation}
The regime of $w\ll 1$ is possible only if $\xi\ll 1$. In this case, $w_1=2\xi$ and Equation~(\ref{eqn:w1}) becomes in the leading order,
\begin{equation}
\label{eqn:wc}
   \frac{dw}{d\ln r}=-\frac{2}{3}\left(w-2\xi\right) \quad \Rightarrow\quad w-2\xi\propto r^{-2/3}. 
\end{equation}

\bibliographystyle{apj}

\end{document}